# Symmetry-breaking-induced giant Stark effect in 2D Janus materials

Jiang-Yu Lu[1], Wu-Yu Chen[1], Lei Li[1], Tao Huang[1], Hui Wan[1,2], Zi-Xuan Yang[1], Gui-Fang Huang[1*], Wangyu Hu[3], Wei-Qing Huang[1*]

[1]Department of Applied Physics, School of Physics and Electronics, Hunan University, Changsha 410082, China
[2]School of Materials and Environmental Engineering, Changsha University, Changsha, 410082, China
[3]College of Materials Science and Engineering, Hunan University, Changsha, 410082, China
*Corresponding Authors: gfhuang@hnu.edu.cn; wqhuang@hnu.edu.cn

**Abstract**: Symmetry breaking generally induce exotic physical properties, particularly for low-dimensional materials. Herein we demonstrate that symmetry breaking induces a giant Stark effect in 2D Janus materials using group IV-V monolayers with a four-atom-layer structure as a model system, which are constructed by Ge and As element substitution of symmetrical SnSb monolayer. A linear giant Stark effect is found in Janus semiconductor monolayers, as verified by the band gap variation up to 134 meV of $Sn_2SbAs$ monolayer, which is 30 times larger than that of SnSb monolayer (4 meV) when the applied electric field is increased from -0.30 to 0.30 V/Å. By considering the induced electronic field, we propose a generalized and effective formula that efficiently determines the band gap variation owing to Stark effect. The calculated results from proposed formula are well agreement with those from DFT-HSE06 functional. The giant Stark effect is originated from the large spatial separation of centers of the conduction band minimum and valence band maximum states of Janus structure due to its intrinsic potential gradient. The wide-range tuning of band gap under electronic field shows potential applications of 2D Janus materials in optoelectronic devices.

Symmetry principles are universally utilized to describe important phenomenological properties of systems by the scientific community and to disclose key mechanisms by simplifying the solutions without tedious and complicated calculations. When the symmetry of a system is broken[1], peculiar physical properties, such as ferromagnetism[2], plasmonic exceptional points[3], ferroelectricity[4] and valley Hall effect[5] will emerge, thus enhancing functions or inducing new-/multi-functionalities. In essence, symmetry breaking is ubiquitous in nature, which has brought up the richness and variety of world. Breaking the symmetry is therefore considered to be an effective strategy to create novel physics and functionality.

Since graphene was isolated utilizing the 'Scotch Tape' and characterized for the first time in 2004, two-dimensional (2D) material research has accelerated exponentially[6-11]. A tremendous amount of 2D materials has been designed theoretically or/and successfully synthesized experimentally, including single-element-based 2D materials (such as silicene[12], germanene[13] and stanene[14]), binary, ternary and multinary 2D compounds[6, 8, 15]. As the artificially synthesized materials, the symmetry of 2D materials would be easily to break by adjusting the process parameters. In fact, doping and substitution have become common methods to break the symmetry of 2D materials. In particular, 2D materials with multi-atom-layer structure (such as transition metal dichalcogenides (TMDs)) naturally have the possibility to intrinsically break the out-of-plane mirror symmetry. One paradigm is the semiconducting Janus MoSSe with one face consisting of Se atoms and the other of S atoms, which was successfully realized independently by two groups in 2017[16, 17]. This breakthrough has inspired a surge of theoretical and experimental research into 2D Janus materials that exhibit novel physical and chemical properties, including a long exciton radiative recombination lifetime, considerable spin Hall conductivities, tunable band gap and electron transport[18, 19]. These unique properties make 2D Janus materials having great potential for applications in clean energy conversion (such as photovoltaics and water splitting[20-22]) and nanoelectronics (such as valleytronics[23]).

Stark effect is the shifting and splitting of electronic energy levels under an external electric field, and is essential for device applications[24-33]. Specifically, the external electric field leads to the redistribution of valence band maximum (VBM) and conduction band minimum (CBM) states in real space, while the potential difference and band mixing between VBM and CBM states will cause the variation of band gap. Various 2D semiconductors exhibit giant Stark effect[34-36], which is originated from the interlayer potential gradient by an external electric field. In this regard, it is interesting that there exists intrinsically an intralayer potential gradient in 2D Janus materials. Naturally, two key issues arise: (1) what is the role of intralayer potential gradient for Stark effect in 2D Janus materials? (2) Under an external electric field, how to accurately describe the effect of potential gradient on Stark effect in 2D Janus materials?

Herein, we systematically explore the Stark effect in 2D Janus materials using a model system of group IV-V monolayers with a four-atom-layer structure. To compare with symmetrical structure, six Janus group IV-V monolayers are constructed based on symmetrical SnSb monolayer by an element substitution strategy. Under an external electric field, SnSb monolayer shows a weak quadratic Stark effect. By contrast, symmetry breaking gives rise to a linear giant Stark effect in 2D Janus materials. By taking the induced electric field into account, we propose a generalized and effective formula that can accurately describe the band gap change under external electric filed, which is confirmed by the band gaps calculated by HSE06 functional. We find that the giant Stark effect can be attributes to the large spatial separation of centers of the conduction band minimum and valence band maximum states of Janus structure owing to its intrinsic potential gradient. These results will provide valuable theoretical guidance for band-engineering.

All the calculations are performed using the first principles theory within the framework of density functional theory (DFT) as implemented in the Vienna ab initio simulation package (VASP) which bases on projector augmented-wave method (PAW)[37-39]. We use the Perdew-Burke-Ernzerhof (PBE) function of the generalized gradient approximation (GGA) given by the exchange potential and the correlation potential to calculate the geometric structure[40]. Considering that the PBE function may underestimate the band gap, the Heyd-Scuseria-Ernzerhof hybrid functional (HSE06) is utilized to calculated the band gap of material[41, 42]. In order to improve the accuracy and efficiency of structural optimization, we sample the Brillouin zone on a gamma-centered k-mesh of 11×11×1 in both geometry optimization and self-consistent calculations of Janus monolayers[43]. In order to avoid the artificial interaction between adjacent layers, the vacuum depth in the Z direction is set to 20 Å in the calculation of monolayers. We set the cut-off energy of the plane-wave basis to be 450 eV, and the convergence tolerance of total energy and the threshold of forces and stress tensor are set to $10^{-6}$ eV and 0.01 eV/Å, respectively. In addition, the crystal structure and charge distribution are visualized through VESTA and VASPKIT[44]. Since the Janus materials are asymmetrical, we add dipole correction to the calculations to reduce the calculation error from the intrinsic dipole moment of materials.

Group IV-V elements are demonstrated to have stable AB monolayer structure with A-B-B-A stacking. Based on the symmetrical SnSb structure, we use a substitution strategy to design Janus group IV-V monolayers, as displayed in **Fig.1**. For example, when the upper Sn atoms are replaced by Ge atoms in SnSb structure, Janus GeSnSb$_2$ monolayer is obtained. By this way, six Janus monolayers (Ge$_2$SbAs, Sn$_2$SbAs, GeSnSb$_2$, GeSnAs$_2$, SbSn-GeAs, and SbGe-SnAs) are generated by using Ge, As, Sn and Sb elements, as shown in **Fig.S1**. Structural and electronic properties of both symmetrical and Janus group IV-V monolayers are calculated and listed in **Table I**. Lattice constants and bond lengths ($l_{\text{Sb-Ge(Sn)}}$, $l_{\text{Ge-Sn}}$, $l_{\text{Sn(Ge)-As(Sb)}}$) are obtained by performing structural optimization. Due to the difference of elements and their combinations, lattice constant $a_0$ and bond length $l$ of seven monolayers are different: the former ranges from 3.95 Å to 4.38 Å, and the latter varies form 2.54 Å to 2.88 Å. The band structures of all monolayers are theoretically calculated using the HSE06 functional and displayed in **Fig.S2**. It is found that the SbSn-GeAs monolayer exhibits a metallic behavior, while other monolayers show semiconducting characteristic with bandgaps of 0.76-1.30 eV. Different from the symmetrical SnSb monolayer, six Janus monolayers have an intrinsic electric polarization owing to the symmetry breaking, and their dipole moment per unit cell are calculated to be about 0.034-0.113 D, as shown in **Table I**.

We next turn to the band structure of Janus group IV-V monolayers, which are shown in **Fig. S2**. Among the semiconductors mentioned above, Ge$_2$SbAs and Sn$_2$SbAs have direct band gaps with the valence band maximum (VBM) and the conduction band minimum (CBM) located at the Γ point in the Brillouin zone (BZ). While the others (GeSnSb$_2$, GeSnAs$_2$, SbGe-SnAs and SnSb monolayers) exhibit indirect band gaps between the VBM at Γ point and CBM at M point. **Figure 2** illuminates the charge density distributions of the VBM states at the Γ point (VBM@Γ), and the CBM states at the Γ and M points (CBM@Γ, CBM@M) of SnSb, SbGe-SnAs and GeSnSb$_2$ monolayers, respectively. Due to its structural symmetry, the charge density distribution in real space are mirror symmetry in the SnSb monolayer (**Fig. 2(a)**). For Janus monolayers, however, the vertical mirror symmetry breaking leads to the asymmetry of the charge density distribution of the VBM and CBM states. For instance, the VBM@Γ of SbGe-SnAs monolayer is dominated by the Sb and Ge atoms, while the CBM@Γ mainly originates from Sn and As atoms (**Fig. 2(b)**). In particular, the VBM@Γ charge density of GeSnSb$_2$ monolayer is totally localized in the lower Sn and Sb atoms, whereas the CBM@Γ only originates from the upper Sb and Ge atoms (**Fig. 2(c)**). Meanwhile, the VBM@M charge density has a vertical mirror asymmetry in the Janus monolayers.

Applying an electric field is an effective method to regulate the electronic properties of 2D materials. Now, we focus on the band gap and band gap variation of six semiconductor monolayers under an out-of-plane electric field ranging from -0.30 V/Å to 0.30 V/Å, as illustrated in **Fig. 3** and **Figs. S3-S8**. As expected, the band gap of symmetrical SnSb monolayer shows a small nonlinear change under the electric field, corresponding to a weak, predominantly quadratic Stark effect (the inset in **Fig. 3(a)**). By contrast, five Janus semiconductor monolayers display a linear giant Stark effect (**Fig. 3(a, b)**). For example, the band gap variation of Janus Sn$_2$SbAs can reach up to 134 meV, which is 30 times larger than that of symmetrical SnSb monolayer (4 meV) when the applied vertical electric field is increased from -0.30 V/Å to 0.30 V/Å. These results demonstrate that symmetry-breaking could induce the giant Stark effect in 2D Janus materials.

Generally, the linear Stark effect can be calculated by a simple formula[45]:

$$E_g(\varepsilon_E) - E_g(0) = e\varepsilon_E(\langle z\rangle_v - \langle z\rangle_c) = e\varepsilon_E d \qquad (1)$$

where $E_g(\varepsilon_E)$ and $E_g(0)$ is the band gap at external electric field $\varepsilon = \varepsilon_E$ and zero field, respectively; $\langle z\rangle_v = \frac{\langle\phi_v|z|\phi_v\rangle}{\langle\phi_v|\phi_v\rangle}$ and $\langle z\rangle_c = \frac{\langle\phi_c|z|\phi_c\rangle}{\langle\phi_c|\phi_c\rangle}$ are the charge center of VBM state and CBM state, respectively; $d$ is the distance between $\langle z\rangle_c$ and $\langle z\rangle_v$. Here, we set $d > 0$ when the center of VBM state is above the center of CBM state in real space when projected on z axis. When Eq. (1) was proposed and used[45, 46], $d$ is always treated as a constant. In fact, $d$ should be a function of the applied electric field strength, because electrons will redistribute under an external electric field. Therefore, Eq. (1) should be rewritten as:

$$E_g(\varepsilon_E) - E_g(0) = e \int_0^{\varepsilon_E} (\langle z \rangle_v - \langle z \rangle_c) d\varepsilon = e \int_0^{\varepsilon_E} d(\varepsilon) d\varepsilon \qquad (2)$$

Eq. (2) is an integral form of Eq. (1), but it still greatly overestimates the band gap variation of Janus 2D materials, as displayed by the dotted line in Fig. 4(a1, a2). Apparently, a correction term is still needed to describe the unusual band gap variation of Janus materials under an electric field. Due to the mirror symmetry-breaking, an intrinsic electric field exists across the Janus materials. When an external electric field is applied, an induced electric field will be appeared in Janus materials due to the redistribution of electrons. Reviewing at Eq. (2), an induced electric field should be added necessarily. The induced electric field, that is, the change of the internal electric field, is in the direction opposite to the external electric field and proportional to the strength of the external electric field, appearing as a partial cancellation of the external electric field. Considering the effect of induced electric field, Eq. (2) should be modified as the following expression:

$$E_g(\varepsilon_E) - E_g(0) = e \int_0^{\varepsilon_E} d(\varepsilon) d(\varepsilon + \varepsilon') = e(1 + C) \int_0^{\varepsilon_E} d(\varepsilon) d\varepsilon \qquad (3)$$

where $\varepsilon' = C\varepsilon$ is the induced electric field, C is the proportional coefficient between the induced electric field and the applied electric field. To calculate the induced electric field, we need to first calculate the induced potential, which can be calculated using the following formulae:

$$P_{ind}(Z, \varepsilon_E) = P(Z, \varepsilon_E) + e\varepsilon_E Z - P(Z, 0) \qquad (4)$$

where $P_{ind}(Z, \varepsilon_E)$ is the planar average induced potential at the position $Z$ under the external electric field $\varepsilon = \varepsilon_E$, $P(Z, \varepsilon_E)$ is the planar average total potential at the position $Z$ under the external electric field $\varepsilon = \varepsilon_E$, $P(Z, 0)$ is the planar average total potential at the position $Z$ at zero bias. Fig. 5 displays the calculated plane average induced potential of symmetrical SnSb and five Janus semiconductor monolayers. One can see that the planar average induced potential is almost linear inside the monolayers, indicating that the induced electric field in the z direction can be regarded as a uniform electric field. We linearly fit the plane average induced potential in the range $z \in [8, 13.5]$, and the obtained slope is the value of the induced electric field. The induced electric field strength of the six semiconductor monolayers is proportional to the applied electric field, and the proportional coefficient C is about -0.94, as given in Table II.

Based on the calculated results above, Eq. (3) is used to simulate the band gaps of six semiconductor monolayers under a vertical electric field, and the calculated results are illustrated by the solid line in Fig. 4(a1, a2). Clearly, the simulation results of Eq. (3) agree well with those of the band-gap calculation of HSE06 functional. It should be pointed out that the simulation error tends to increase with the electric field, mainly due to the error accumulation caused by the insufficiently small step size limited by the calculation cost.

By now, we can reveal the reason why the symmetry breaking enhances the Stark effect. The difference in band gap variation between the six semiconductor monolayers (Fig. 3(b)) is mainly caused by the difference in $d(\varepsilon)$. For Janus monolayers, the symmetry breaking causes the centers of VBM and CBM to spatially separate at zero bias due to the intrinsic potential gradient, while those of symmetric monolayer inevitably coincide due to its symmetry [Fig. 2]. That is to say, the Janus monolayers exhibit a $d$ value larger than zero, but the symmetric monolayer registers a $d$ value of zero at zero bias. When applying an external electric field, the $d$ variation of Janus monolayers is smaller than that of symmetric SnSb monolayer. As shown in Fig. 4(b), the $d$ variation is small in five Janus monolayers; while the $d$ varies linearly in symmetric SnSb monolayer with the electric field and changes from positive to negative. Thus, the $d$ of five Janus monolayers always maintains at large values (the absolute value of $d$ is bigger than 2.5 for Janus monolayers except for SbGe-SnAs monolayer, Fig. 4(b)), whereas that of symmetric structure will keep in small values (the absolute value of $d$ is smaller than 0.5) in the external electric field strength range studied. From Eq. (3), the greater the value of $d$, the larger the variation in the band gap. Therefore, for a symmetric structure, the small $d$ and its variation from positive to negative will result in a small quadratic change of band gap with an applied electric field; while the big $d$ and its small, linear variation without the alternatively positive and negative change will give rise to a large variation of band gap (i.e., the giant Stark effect) in Janus monolayers.

In summary, we have shown that six Janus group IV-V monolayers can be constructed by using Ge, As, Sn and Sb elements based on the symmetrical SnSb structure, and the giant Stark effect appears in these semiconductor Janus monolayers when subject to an external field. The symmetry-breaking-induced giant Stark effect can be attributed to the large, natural spatial separation of centers of VBM and CBM of Janus group IV-V monolayers. We also propose a general formula for band gap variation due to giant Stark effect by considering the effect of induced electronic field under an external field. The calculated results from proposed formula are well agreement with those from DFT-HSE06 functional. These results imply that the band gap of Janus materials can be readily increased or decreased by changing the direction of external electronic field. Our work reveals the physical mechanisms of the giant Stark effect and provides a powerful tool for energy band engineering in Janus materials.

See the supplementary material for the crystal structures of Janus monolayers, projected band structures and their corresponding changes under external electric field of $Ge_2SbAs$, $Sn_2SbAs$, $GeSnSb_2$, $GeSnAs_2$, SbSn-GeAs, SbGe-SnAs and SnSb monolayers.

This work was supported by the National Natural Science Foundation of China (No. 52172088) and Natural Science Foundation of Hunan Province (No. 2021JJ30112).

## DATA AVAILABILITY

The data that support the findings of this study are available from the corresponding author upon reasonable request.

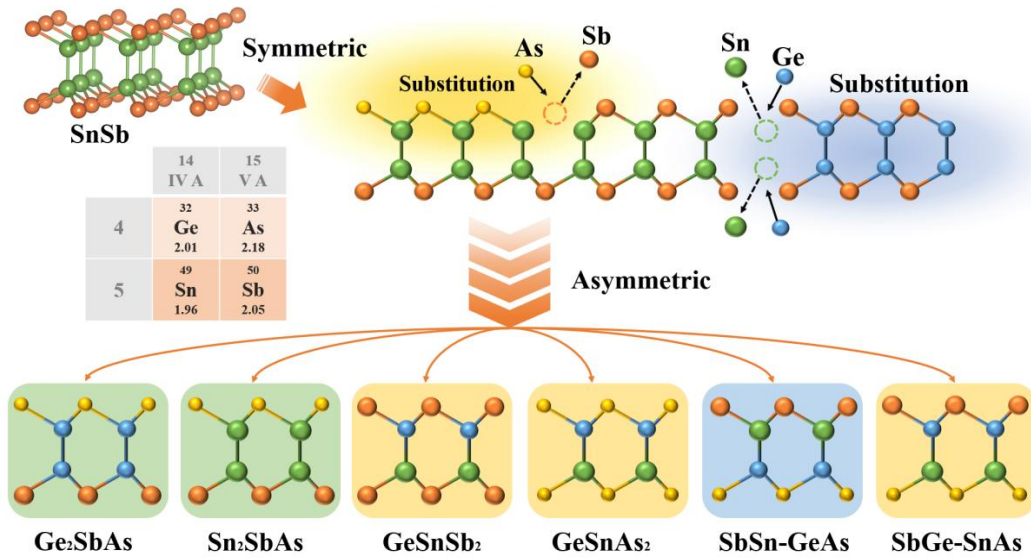

**FIG. 1.** Optimized atomic structures of SnSb and its Janus monolayers. The green, yellow and blue backgrounds respectively represent direct, indirect and zero band gap.

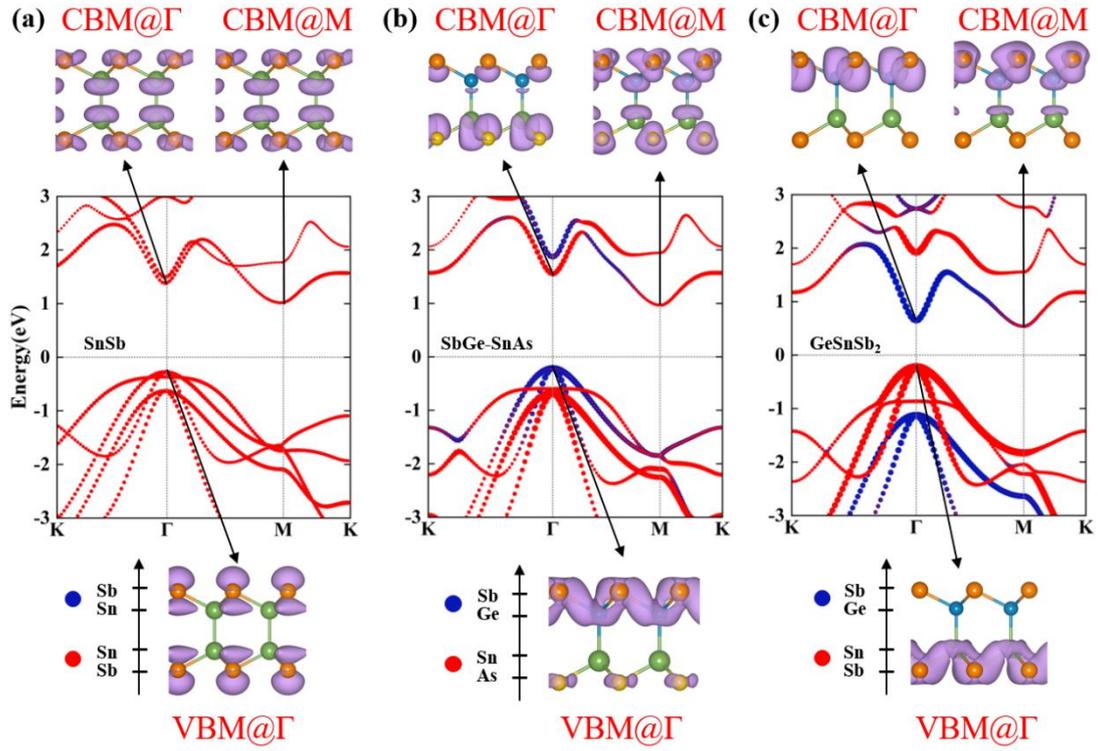

**FIG. 2**. Projected band structures calculated by HSE06 and charge density for VBM@Γ, CBM@Γ and CBM@M states of (a) SnSb, (b) SbGe-SnAs and (c) GeSnSb$_2$ monolayer. The fermi levels are set to zero.

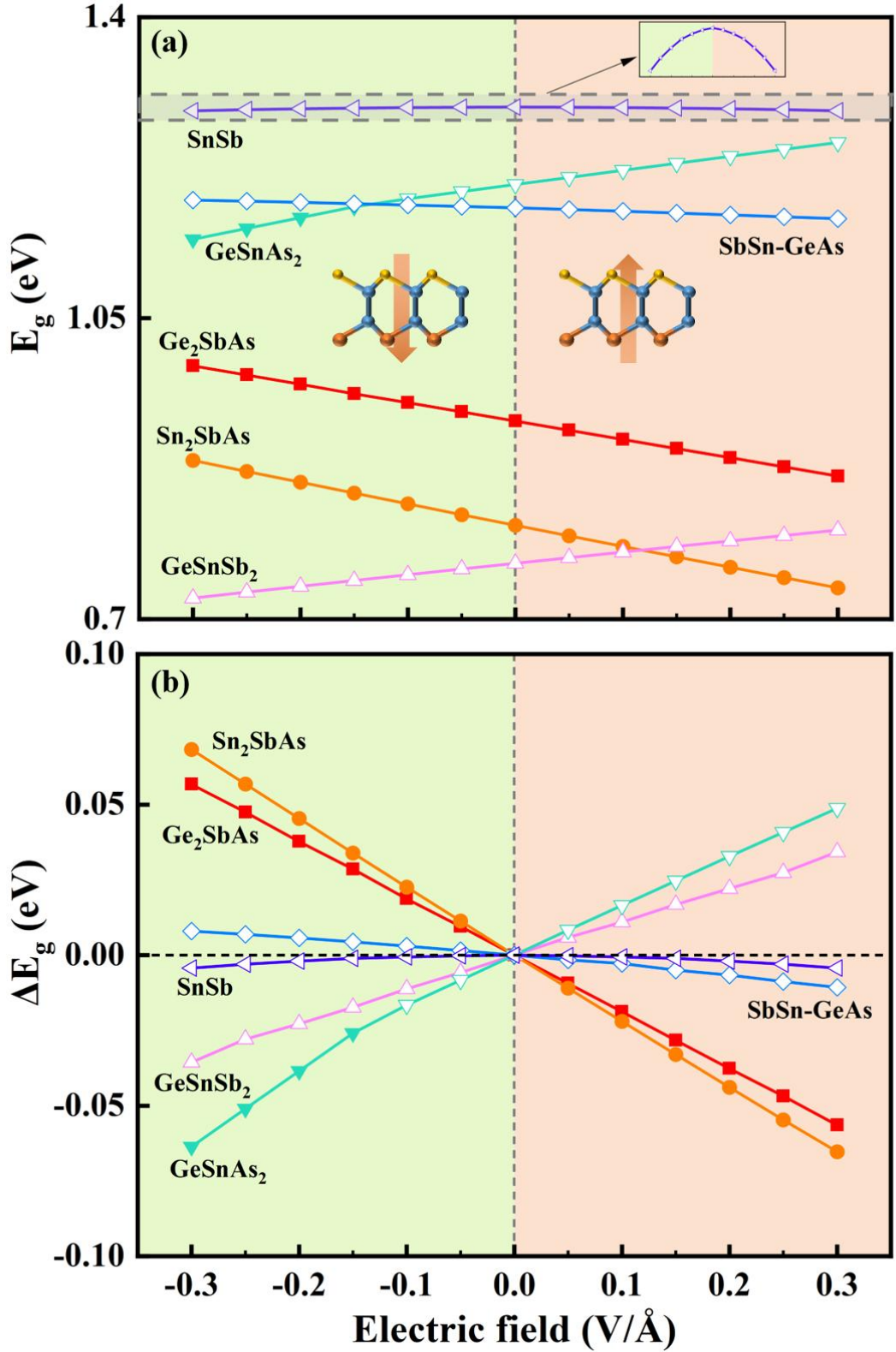

**FIG. 3.** Variation trend of the (a) band gap $E_g$ and (b) band gap variation $\Delta E_g$ with electric field of six semiconductors. The direct (indirect) band gap is marked with solid (hollow) points.

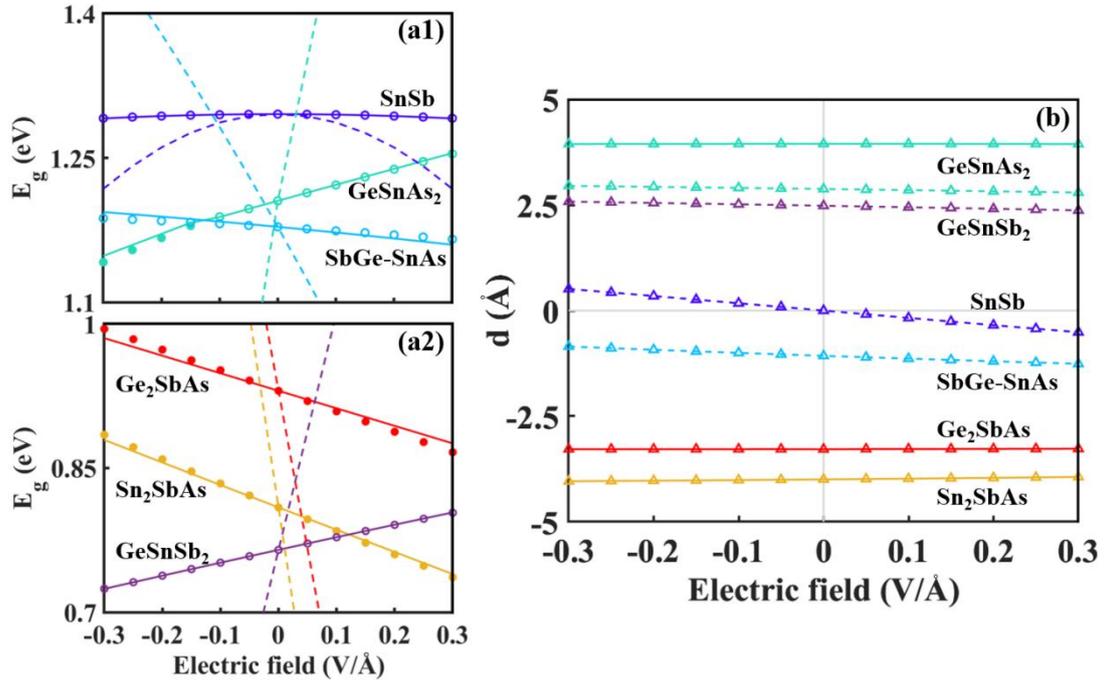

**FIG. 4.** Variation trend of $E_g$ with electric field of (a1) SnSb, SbGe-SnAs, GeSnAs$_2$ monolayers and (a2) Ge$_2$SbAs, Sn$_2$SbAs, GeSnSb$_2$ monolayers. The dashed and solid line respectively represent calculated by eq. (2) and eq. (3). (b) Variation trend of d with electric field. The solid (dashed) line represents the distance between VBM@Γ and CBM@Γ (VBM@Γ and CBM@M).

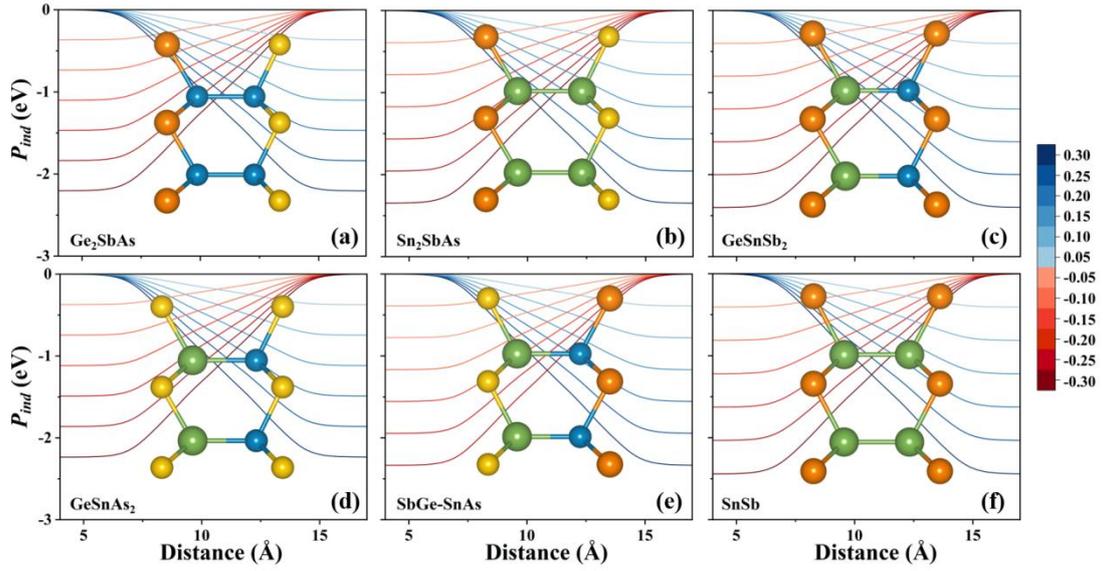

**FIG. 5.** The planar average induced potential of (a) Ge$_2$SbAs, (b) Sn$_2$SbAs, (c) GeSnSb$_2$, (d) GeSnAs$_2$, (e) SbGe-SnAs, (f) SnSb monolayer under different electric field strengths from -0.30 to 0.30 V/Å.

**TABLE I.** Calculated lattice parameter $a_0$ (Å), bond length $l$ (Å), dipole moment $d_\mu$ (D), band gap $E_g$ (eV) of each monolayer.

| Monolayers | $a_0$ (Å) | $l$ (Å) | | | $d_\mu$ (D) | $E_g$ (eV) |
|---|---|---|---|---|---|---|
| | | Sb(As)-Ge(Sn) | Ge-Sn | Sn(Ge)-Sb(As) | | |
| Ge$_2$SbAs | 3.96 | 2.64 | 2.49 | 2.54 | 0.086 | 1.04 |
| Sn$_2$SbAs | 4.23 | 2.83 | 2.88 | 2.72 | 0.079 | 0.94 |
| GeSnSb$_2$ | 4.25 | 2.83 | 2.68 | 2.74 | 0.036 | 0.76 |
| GeSnAs$_2$ | 3.95 | 2.54 | 2.69 | 2.62 | 0.034 | 1.21 |
| AsGe-SnSb | 4.09 | 2.80 | 2.68 | 2.59 | 0.079 | 0 |
| SbGe-SnAs | 4.10 | 2.69 | 2.68 | 2.67 | 0.113 | 1.20 |
| SnSb | 4.38 | 2.87 | 2.87 | 2.87 | 0 | 1.30 |

**TABLE II.** The proportional coefficient C of induced electric field and applied electric field of six semiconductor monolayers.

| Monolayers | Ge$_2$SbAs | Sn$_2$SbAs | GeSnSb$_2$ | GeSnAs$_2$ | SbGe-SnAs | SnSb |
|---|---|---|---|---|---|---|
| C | -0.944 | -0.942 | -0.947 | -0.942 | -0.947 | -0.942 |